\documentclass[12pt]{article}
\usepackage{graphicx}
\usepackage{amsfonts}
\usepackage{amscd}
\usepackage{latexsym}
\usepackage{epsf}

\newcommand{\vs}[1]{\vspace*{#1\linewidth}}
\newcommand{\hs}[1]{\hspace*{#1\linewidth}}
\newcommand{\be}{\begin{equation}}
\newcommand{\ee}{\end{equation}}
\newcommand{\bea}{\begin{eqnarray*}}
\newcommand{\eea}{\end{eqnarray*}}
\newcommand{\barr}{\begin{array}}
\newcommand{\earr}{\end{array}}
\newcommand{\bal}{\begin{align*}}
\newcommand{\eal}{\end{align*}}
\newcommand{\bi}{\begin{itemize}}
\newcommand{\ei}{\end{itemize}}

\newcommand{\nn}{\nonumber}
\newcommand{\fr}{\frac}
\newcommand{\pd}{\partial}
\newcommand{\al}{ \alpha }
\newcommand{\la}{ \lambda }

\newcommand{\vp}{\varphi }
\newcommand{\vep}{\varepsilon }

\newcommand{\ra}{\rightarrow}

\usepackage[latin9]{inputenc}
\usepackage{textcomp}
\usepackage{amsmath}
\usepackage{multirow}
\usepackage{color}
 
\newcommand {\ol}[1]{\overline{#1 }}

\newcommand {\ga} {\gamma}

\newcommand {\de} {\delta}  
\begin{document}
\title{\bf   On a local formalism for time evolution of dynamical systems  }
\author{{\sc I.~V.~Drozdov}\thanks{Corresponding author, e-mail: drosdow@uni-koblenz.de}}
\date{ }
\maketitle
\begin{center} 
\small   Koblenz University, 
\small   Germany \\ 
\date{30 December 2014 }
\end{center} 
\begin{abstract}
 The formalism  of local maximization for entropy gradient producing the evolution and dynamical equations for closed systems. It eliminates the
 inconsistency between the reversibilty of time in dynamical equations and the strict direction of irreversible evolution for complex systems, causality contradictions and 
ambiguity of time flow  in different systems.  Independently it leads to basic principles of special relativity.
\end{abstract}
\section{Introduction}
\subsection{ Evolution \&  Dynamics }
The conventional  understanding
of {\it dynamics} of any processes in the nature is the {\it time dependance } of  state of a system.

Here, the primary concept of any  formulation of {\it dynamics} is the  {\it state}.
   
 A {\it state}  is determined by the complete set of attributes - degrees of freedom (DoF) of the system.

 As a {\it process} we understood the change of state in the {\it time} as a variable parameter.
   The {\it dynamics} is understood as a set of states for different values of the {\it time} variable, 
 which are ordered by increasing values of time. \\
 It results as a solution of {\it dynamical equations} 

The state of a complex system
is defined as a collectivity of single systems - components. 
 For  complex systems with a sufficiently high number of equal (or similar) components 
the description of state is formulated in terms of statisical functions ( {\it statistical systems} ), such 
as statistical sums, mean values, entropy, dispersion, fluctuations and further concepts derived from them. 
 
 The dymanics (or {\it evolution} ) of its statistical states is formulated in terms of the same (universal) time, as for each component.

\subsection{Dynamical equations - short review of conventional approaches}
\subsubsection{ Lagrangian  formalism  }
 The minimization condition for the {\it action functional}
$$
S=\int\limits_{t_i}^{t_f} { \cal L} \left[    q_i , \dot{q_i} , t   \right]    dt  \ra min
$$
 produces the {\it Euler-Lagrange } dynamical equations
$$
   \fr{d}{dt} \fr{\pd}{\pd \dot{q}_i } {\cal   L}      =    \fr{\pd}{\pd  q_i } \cal{L} 
$$
for DoF's  $  q_i $, 
which are of second order in external time $t$  and can be therefore invariant under time reversion $t\ra -t$ \cite{landau_mech}\\

  The  Lagrangian  formalism reveals two obvious shortcomings : 
\begin{itemize}
\item {\bf causality\\}
- the main drawback of this approach is its  {\it global formulation}. 

 The {\it local } dynamical equations  are obtained from the {\it integral} over
 the total  evolution time. 
it means: in order to define its {\it local} behaviour -
{ \bf the at a given time  in a given state - state space point}- 
the system should 'know' all possible states in all times in the past and the future  $\ra$
thus the {\bf causality}   is  {\it a priori} disregarded in this formulation
\item the {\bf  time reversibility} does not reproduce the irreversible evolution of complex systems
\end{itemize}%

\subsubsection{Hamiltonian  formalism }
 The state space is determined by the double set of DoF's $q_i, p_i$ by means of local scalar function 
$$
H=H(q_i, p_i),-
$$
the Hamiltonian function - the key object generating the evolution in the state space.

The time dependence - dynamics-  is introduced externally by imposing the dynamical equations
$$ \fr{dq}{ dt}=H_p;\ \    \fr{d p}{ dt}=-H_q$$
- the {\it Hamilton equations.}
The following  main disadvantages of this approach should be mentioned here are
\begin{itemize}
\item 
redundant degrees of freedom (2n instead of n), although $p_i$ are generally connected with time derivatives of $q_i$;
\item time is an extrinsic parameter -  the  time scale of the dynamics is not an intrinsic feature of the system and can be chosen arbitrary. 
\end{itemize}
\subsubsection{ General shortcomings of conventional formulations}
\begin{itemize}
\item
  {\bf  time irreversibilty  is not regarded }, that causes inconsistensies with generalisation for complex ststistical systems;
\item
since the {\bf causality is generally not assured,} it must be demanded externally by conditions for space-time interval (special relativity) \cite{landau_field};
it causes an ambiguity for definitions of time flow - time is subjected to Lorentz transformations  (especially resulting in relativity of simultaneity)
\end{itemize}
Summarizing the above-mentioned  weak points, several  reasons for a partial revision of this historically formated views are in particular the following:
  \subsubsection{   Time direction problem}
 According    to the conventional belief, the time (as well as the space) is an {\it a priori} existent 
 parameter, external for a considered system 
among all subsystems and observer ( system  of observation ). 
 The  canonical set of equations, formulated in terms of such time, can possess a symmetry property, 
 regarding the time (revesibility), e.g. if the equations are of an even order (as for the case of newtonian mechanics).

  On the other hand,  a great macro-system consisting of a plurality of elemetrary time - reversible systems 
never possesses a reversibility. Instead of this, the evolution of a macro-system follws the {\it second law of thermodynamics}, according to which the entropy of the closed system, regarding the increasing time,  cannot decrease. 

 Thus a conventional (e.g. the Newtonian or Lagrangian) formulation   leads to
the general  inconsistency between the reversibilty of time in dynamic equations and the strict direction 
of irreversible evolution of complex systems consisting of a plurality of elementary  systems subjected to the time-reversible dynamic equations.

 In other words, the time as an evolution parameter for elementary (reversible ) systems and the global
 time of (irreversibe) macroscopic evolution is apparently not one and the same.
  This evidence  suggests probably, that the priority or {\it hierarchy  } of basic principles for elementary systems and for macro-systems should be re-ordered, what (at least formally) eliminates this contradiction.
%

 \subsubsection{  Causality problem }
 
 The causality is the time-ordering of states according to the  sequence {\it cause - result } 
 for elementary systems as well as for complex systems.

  The relativity of time, resulting from the relativity theory, brings  a certain ambiguity in the
 interpretation of the resulting dynamics, even in the classical level. 
Together with the time revesibility mentioned above;
as well as a demand to distinguish two cases (e.g. advanced-retarded solution)
additionaly arising. 

 Furthemore in a quantum level, harder contradictions occur (e.g. Schroedinger's cat, EPS-paradox)
 
    Another issue for troubles with causality, especially in the classical mechanics
 and field theories, is  the conventional formulation of component dynamics in terms of Lagrangian formalism.
 As it was mentioned above the main drawback of this approach (fatal for causality) is the sufficiently {\it global formulation} - the key  object, producing the local dynamic equations ( action functional) - is the integral over the total configuration space and the evolution time. 
   In simple words it means, in order to define its behaviour at the certain time in the certain configuration state, the system shoud "know" its behaviour in all possible states
   in all times in the past and future. Thus, the causality is disregarded in the formulation from the beginning on.

  \subsection{ Re-deinition of time  }
 Time flows differently  in different systems at different conditions. The ultimate solution for this problem  would be to suppose that the "time" does not exists in general as a physically well-defined measure.   

 The possible revision of the time concept  can be a considering of the time
{\it defined-as-measured} instead of assuming the existence of time {\it a priori} (like the definition of a wave velocity \cite{local}):

   Observe the evolution of some subsystem S (as "system") of the total closed system.
 The "measurement of time" assigned to the each state of S means,  being compared with the state of another subsystem C (as "clock") corresponding to the same {\it global} state of the total system. (Fig.1)

   The "time direction" means, the states of the system C are ordered with respect to the entropy of the total system.
And the "time  evolution" of the subsystem S is the ordering of its states with respect to the states  of subsystem C.

 \subsection{ Entropy maximization principle }

 Finally, the ordering of the {\it global states } of the total closed system  follows the increase of its entropy. 

Thus, the basic postulate, generating the ordering (interpreted as evolution) of a closed sytem among all its subsystems is  the ordering by entropy, or {\it second law of thermodynamics} applied inversely.
  
 Finally, this principle can be  re-interpreted in a {\it local formulation}:
the "trajectory" of the total system in the state space obeys the condition of the entropy maximization,  the  state change occurs in the direction of maximal increasing entropy.  
  Thus, this principle can be called {\it  entropy-gradient-maximization}.

 \subsection{ Local formulation}
 Summarizing the suggestions sketched above, the "dynamics" of the system with arbitrary DoF's results from following principles: 
  \bi
	\item
  the formulation is deterministic: all possible states and the structure of the state space for a closed system are determined by values of its DoF's
	\item 
the trajectories are determined by initial points and the entropy gradient maximization principle with additional conditions
	\item 
 the topological structure of the state space is determined by paths (trajectories) between points,
 where the starting points corresponds to the {\it cause }is the initial condition of path; the causality is then uniquely fixed;
 	\item
 the "time dynamics" is the ordering of the points of trajectory with respect to substates of a chosen subsystem  (clock system);
\ei
 
 The mathematical formalisation of these principles is developed in the next sections    below, proceeding from the example of euclidean space with the euclidean norm,  
further generalization  on the case of an arbitrary state space with an arbitrary ergodicity condition is formulated;

The resultung dynamical equations are considered first for a sytem with two DoF's, with the resulting corollaries for time and causality restrictions.

The further generalization for a system with 3 DoF  points the common way for further generalization of the formalism for more DoF's. In particullary, the issue of the time ambiguity or "relativity" is explained.
In the  last  section the argumentations and resulting implications are concluded.
\section{Postulates of the formalism}

\subsection{  {\it  Entropy-gradient-maximization} principle  of  evolution   }     
 The problem of {\bf  time direction and causality } can be eliminated by rearranging the
 hierarchy of key principles: 
\begin{itemize}
	\item
the primary (generating) principle is the  {\it second law of thermodynamics}  -  {\bf entropy maximization} - the ordering of {\it global states } of a total closed system  follows the increase of its entropy(in conventional approaches it should be fulfilled in a secondary priority); 
	\item
this formulation should be {\it local }:  the state evolution in each point of the  state space occurs in the direction of maximal increasing entropy;
	 \item
 each solution for the 'trajectory' of the total system in the state space obeys therefore the condition of the entropy maximization, which is defined as {\bf causality}; 
\end{itemize}

\subsection{ Definition of time   'as-measured'  }    
 Observe the evolution of some subsystem S (as 'system') of the total closed system (Fig.1)
\begin{figure}
\centering\includegraphics[scale=0.52]{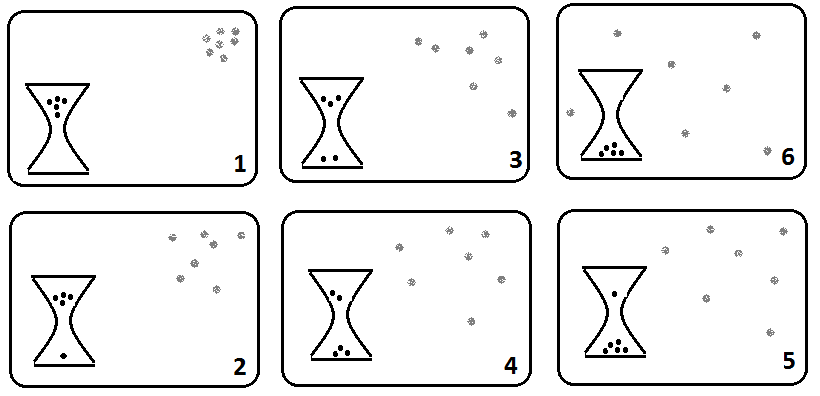}\\
\caption{\hspace*{0.5cm}\vspace*{0.5cm}{\small\bf Ordering of states of a closed system by increasing entropy} }
\end{figure}Measurement procedure\ \
\begin{itemize} 
\item
 States of S are compared with states of a subsystem C ('clock' ): \\
  each state of S corresponds to the state of a subsystem C in the same {\it global} state of the total system
\item 
'time  evolution' of the subsystem S is the ordering of its states with respect to the states of subsystem C
\item 
    'time direction' means that states of the system C are ordered with respect to the entropy increasing of the total system
\end{itemize}

\subsection{ Introducing example of gradient maximization - gradient of a scalar field  }
  Consider the scalar field $\vp(x_i) $, defined on the  n-dimensional 
euclidean space $\vec{x}= \{ x_i \} $

 The property  of  {\it gradient} vector   $\nabla\vp(x_i) $ 
$$
\nabla\vp(x_i) := \left\{\fr{\pd}{\pd x_i}\vp   \right\}
$$
is, it points in the direction of maximal increasing of $\vp$ \cite{math}.
 It can be shown as follows:
   Let the point vector $\ol{x_i}$ gets a variation $x_i$ 
 and one seeks its components $x_i$  such that  the variation of $\vp$ in the point $ \ol{x_i} $ maximizes
$$ \de \vp(\ol{x_i}+ x_i)  \ra  max _{x_i} $$
 To this end it is enough to consider the linear part of the variation
 as a function of the variation components
$$   \de \vp(\ol{x_i}+ x_i) =  \left. \fr{\pd }{\pd x_i} \vp(x_i) \right|_{\ol{x_i}}  x_i :=  \left. \nabla\vp  \right|_{\ol{x_i}} x_i  , $$

  which should be maximized with the normalization condition:
$$|\vec{x} |^2=  \sum\limits_{i=0}^n x_i^2  = c^2 =\mbox{  const }$$
  The solution for maximum condition using the Lagrange multiplier method is:

i)\ \   vanishing of first derivatives ( subscript $x_i$ denotes the partial derivative $\fr{\pd }{\pd x_i}$):
$$
\left\{     \de \vp  \right\}_{x_i} =\left\{   \left. \nabla\vp(x_i)  \right|_{\ol{x_i}} x_i  +   \la\left(  \sum\limits_{i=0}^n x_i^2  - c^2 \right)  \right\}_{x_i}  =0
$$
(n equations, $\la$ is a Lagrange multiplier ) \\

ii)\ \ 
non-positivity of the second derivative matrix   $$ \{ \de \vp \}_{x_i x_k}  = \left\{   \left. \nabla\vp  \right|_{\ol{x_i}} x_i  +   \la\left(  \sum\limits_{i=0}^n x_i^2  - c^2 \right)  \right\}_{x_i x_k}$$

  The first requirement (i) provides
$$
   \fr{\pd }{\pd x_i} \vp + 2\la x_i:= \vp_{x_i}+ 2\la x_i=0 , \ \ \    \vp_{x_i}= -  2\la x_i, \ \ i=\ol{0 ... n}
$$
 squared and summarized over $i$ and taking into account the additional (normalization) condition :
$$
\la^2 =     \fr{1}{4c^2}\sum\limits_{i=0}^n \vp_{x_i}^2  ;\ \   \la =    \pm \fr{1}{2c}\sqrt{  \sum\limits_{i=0}^n \vp_{x_i}^2 }
$$
and the second  requirement (ii)  fixes the "+" sign of $\la$:
$$  \vp_{x_i x_k}  = -2  \la \de_{i k }$$, 
what results in the final solution:
 $$ {  x_i = \fr{\vp_{x_i} }{2\la}  } = c \fr{\vp_{x_i} }{  \sqrt{  \sum\limits_{i=0}^n \vp_{x_i}^2 }  } = \fr{ c\nabla\vp(\ol{x_i}) }{  \sqrt{  \sum\limits_{i=0}^n \vp_{x_i}^2 }  }$$
\section{ Formulation of the {\it entropy-gradient-maximization}}
{
\subsection{General       Formulation  -  state ordering    }
    Consider the total  ({\it closed })  system with $n+1$ degrees of freedom (DoF).
$$
q_i=\{     q_1, q_2, ... , q_n \} \mbox{ and } \tau 
$$
 which define the state of system completely (system defined by  $\{ q_i, \tau \}$ is a {\it closed }  system).

The states  $\{ q_i, \tau \}$ are ordered with respect to the  the increasing values of scalar function $S(q_i, \tau)$ -entropy, which can be interpreted as a scalar field on the space $\{ q_i, \tau \}$

 \subsubsection{\hs{.1}      {\large \textbf{  Choice   of   time        }}      }
    The values of the DoF $\tau $   are ordered  so that:
 $$  S(q_i, \tau_1)  \le  S(q_i, \tau_2) \le ...\le S(q_i, \tau_n)  \le S(q_i, \tau_{n+1} \le\ ...)   $$ 
$$
\mbox{ for }   \tau_1 <\tau_2 < \ ...\ < \tau_n< \tau_{n+1} < ...$$
  Form the values  $ \tau_1 <\tau_2 < \ ...\ < \tau_n< \tau_{n+1} $ a continuum, is $ S(q_i, \tau) $ a monotonic non-decreasing  function of $\tau$:
\be
\fr{\pd }{\pd \tau } S := S_\tau \ge 0  \mbox{  (time-eligibility condition for $\tau$)   }
\label{time_eligibility}
\ee

\subsubsection{ First-order-gradient  formalism} 

 Consider the first order variation of the  entropy  $ S(q_i, \tau)$  in the point $ \{  \ol{q_i}, \ol{\tau} \} $:

\be
     \de S(    \ol{q_i}, \ol{\tau} )   =  S_\tau  \tau + \sum\limits_i   S_{q_i} q_i
\ee
 (values of partial derivatives are taken in the point  $ \{  \ol{q_i}, \ol{\tau} \} $ )

 The change of state occurs in the direction, where  the  entropy $\de S(    \ol{q_i}, \ol{\tau} ) $  variation maximizes; 
 with the  $ q_i, \tau $ - variation vector obeying the additional condition:
\be
   h(  q_i, \tau )=\vep_0
\label{ergodicity}
\ee 
- the {\it ergodicity} condition, which means usually some conservation law, e.g. for the
total energy (a simple euclidean norm cannot be used, since the space $ (q_i, \tau)$ is not euclidean).

  It provides the equations:
i)
\bea
&& \de S_\tau= 0\\
&& \de S_{q_i}=0
\label{trajectory}
\eea
 trajectory direction

 for first derivartives and
ii)
\be 
\left[   \barr {ll}    \de S_{\tau\tau}   &     \de S_{\tau\ q_i}     \\
					\de S_{ q_k \tau} &	\de S_{q_i\ q_k}
\earr\right]     \left \{  \Large \le 0\right\} \mbox{ matrix $[n+1 \times n+1] $ is non-positive}
\label{causality}
\ee 
- causality condition
\section{ Example of the 2 DoF system }

  Consider the system possessing 2 DoF's denoted $q,\tau$ with the entropy $ S(q,\tau)$ and 
the ergodicity condition $h(  q, \tau )=\vep_0$
and apply the conditions i) - ii) mentioned above.

 The derivatives can be obtained avoiding the procedure of Lagrange multiplier for the sake of lucidity; 
it provides for $ \de S=\de S(q,\tau)$:

\bea
&& d  \de S =  S_q d q+ S_\tau d\tau \\
&&d h(q,\tau) =h_q d q  + h_\tau d\tau  = 0
\eea

from the latter we have: 
\be  \dot{q}  := \fr{dq}{d\tau} = -\fr{h_\tau}{h_q}= -\fr{S_\tau}{S_q},  (\mbox{ and }\tau_q: = \fr{d\tau}{dq}=-\fr{h_q} {h_\tau} \mbox{ respectively} )
\label{dynamic1ord}
\ee

substituted into the former and with the condition i) applied  :
\begin{eqnarray*}
 && \fr{ d\de S}{d\tau } =S_q \dot{q}+ S_\tau =-\fr{h_\tau}{h_q} S_q + S_\tau=0\\
&&  \fr{ d  \de S}{d\tau } =     S_q + S_\tau \tau_q =   S_q - \fr{h_q}{h_\tau} S_\tau = 0,
\end{eqnarray*}
 two equations which are the same.
Thus the evolution trajectory in $q,\tau$ is governed by a {\bf first-order equation }\\

 For example, the system obeys the ergodicity condition
\be
 h(x,t)=\vep_0
\label{mech_energy}
\ee 
-some kind of energy conservation with the $x$-1D space co-ordinate, $t$
value of ordered states of a clock. It results from the (\ref{dynamic1ord})
$$  -\dot{x} h_x= h_t $$
 and the usual construction $$ h(x,t) = T(\dot{x}) +U(x) = \fr{m}{2} \dot{x}^2+U(x)$$
 leads to $$ -\dot{x}U_x=\dot{x}  m\ddot{x}$$
 providing the dynamical (newtonian) equation and the indentity. It indicates, that the 
time reversibility origins from the essential constraint, contained in the
 conventional form of energy (\ref{mech_energy}). Additionaly, it imposes a requirement
  for special form of entropy by (\ref{dynamic1ord}).
 
 A detailed analysis of conventional dynamical systems is performed in the forthcoming 
 investigation \cite{hamiltonian}.\\

 The second condition results in the matrix
\be
\pd^2 S=\left[   \barr {ll}    \de S_{\tau\tau}   &     \de S_{\tau\ q}     \\
					\de S_{\tau q} &	\de S_{q\ q}
\earr\right]  = -S_\tau \fr{ \al_q- \al \al_\tau}{\al^2} \left[   \barr {rr}   1   &     -\al     \\
					  -\al   &	\al^2
\earr\right]
\ee
 with the notation $\al:=  \fr{h_q}{h_\tau} =   \fr{S_q}{S_\tau}$

Since the matrix is  explicitely non-negative, the condition results in:
$$
{ \al_q- \al \al_\tau}  >0\ \    \mbox{  or  } \ \  \fr{   \dot{q} h_{q\tau}   +h_{\tau\tau}}{   \dot{q}h_{q\tau} - \dot{q}^2 h_{qq}   } < 1,\ \ \dot{q}= -\fr{h_\tau}{h_q} =   -\fr{S_\tau}{S_q}
$$
 Thus, an existence of finite rate of DoF (corresponing to {\bf finite velocity } for relativistic causality) is a direct corollary of the formalism.
\section{ Example of the 3 DoF system }
\subsection{ Application of the formalism}
Now, let the system possess 3 DoF's    $\{ q, r, \tau \}$ 

 The local entropy variation
\be \de S=  S_q {q} +S_r  {r} + S_\tau \tau 
\label{ent_var}\ee
minimizes with the ergodicity 
\be
h(q, r, \tau)=\vep
\label{ergod}
\ee

The  condition i) (\ref{trajectory} ) results in the set of conditions
\bea
&&    S_q \dot{q} +S_r  \dot{r} + S\tau = 0 \\
&&    S_q {q_r} +S_\tau   {\tau_r} + S_r = 0  \\
&&     S_q +S_\tau \tau_q + S_r r_q = 0 \\
&&   h_q \dot{q} +h_r  \dot{r} + h_\tau = 0   \\
&&    h_q {q}_r +h_r  + h_\tau \tau_r = 0 \\
&&   h_q  +h_r  {r_q} + h_\tau \tau_q = 0  \
\eea
 where only two are independent of each other.

 The system ist formally a linear system of 6 inhomogeneous 
equations:
\be
 \left[   \barr{llllll}  
				S_q & S_r  &  &  & &         \\      
				h_q & h_r   &  &  & &          \\      
					&    & S_q & S_\tau & &          \\     
 					&    & h_q & h_\tau & &          \\    
  					&    &  &  & S_\tau & S_r         \\   
					&    &  &  & h_\tau& h_r                        
\earr \right ]      \left[   \barr{l}    \dot{q}  \\   \dot{r}  \\   q_r  \\   \tau_r=1/\dot{r}  \\   \tau_q=1/\dot{q}  \\   r_q=1/q_r  \\       \earr \right ] =  - \left[   \barr{l} S_\tau\\ h_\tau \\   S_r\\ h_r \\S_q\\ h_q \earr \right ] 
\ee  

with the solution:
\be
\dot{q}=-\al := -\fr{D_q}{D_\tau },\  \ \dot{r}=-\beta:= -\fr{D_r}{D_\tau},\ \  \ {q_r}=\ga :=  \fr{D_q}{D_r}
\ee
where the determinants are:
\be   D_q=  \left|  \barr{ll}  S_\tau  & S_r \\ h_\tau  &  h_r   \earr  \right|,  D_r=  \left|  \barr{ll}  S_q  & S_\tau \\ h_q  &  h_\tau   \earr  \right|,  D_\tau=  \left|  \barr{ll}  S_q  & S_r \\ h_q  &  h_r   \earr  \right|,       
  \ee

 The trajectory of the system in the state space $\{   q, r, \tau  \}$ is defined by two equations:
  \be 
	\fr{d q}{ q(\tau)} = \fr{d r}{ r(\tau)} = \fr{d\tau}{\tau } 
  \ee
 where $   q, r, \tau  $  are solutions of the system

\bea
 S_q(q, r, \tau)+S_\tau (q, r, \tau)\fr{1}{\ga}+S_r(q, r, \tau)  \fr{1}{\al} &=&0\nn\\
h(q, r, \tau) &=& \vep
\eea

   The causality matrix ii) consists of 
\bea
 \de S_{\tau \tau} = - S_q \al_\tau -S_r \beta_\tau     & \de S_{qq} = - S_r (1/\al)_q -S_\tau (1/\ga)_q   &  \de S_{rr} = - S_q \ga_r  -S_\tau  (1/\beta)_r\nn\\
 \de S_{q\tau} = - S_q \al_q  -S_r \beta_q      & \de S_{qr} = - S_r(1/\al)_r  -S_\tau  (1/\ga)_r  & \de S_{r\tau} = - S_q \ga_\tau -S_\tau (1/\beta)_\tau\nn
\eea 
 Thus the causality condition consists of 3 inequalities for 3 determinants according to the Sylvester-Jacobi criterion of non-positivity (and generally n inequalities for n degress of freedom).

\subsection{ The  "time  transformation"} 

   In the (\ref{ent_var},\ref{ergod} )  the DoF $\tau$ is chosen to be a "clock" subsystem.

    Since all DoF's   $(q, r, \tau)$  are equivalent in sense  {\it pari passu } to be chosen as a "clock"
  e.g. the DoF $r$  can be considered as a "time " as well.

    Suppose furthermore, the DoF's $\tau$ and $r$ are equal to be chosen as a 'clock':
\be
S_\tau =S_r
\ee
  then
$$
 \dot{q}= -\fr{S\tau}{ S_q} (\dot{r}+1)
$$
 and the same equation reads also
$$
S_\tau \left( \fr{1}{\dot{q}}+\fr{1}{q_r}  \right) +S_q=0
$$
 that means 
\be
\dot{q}= \fr{-S_\tau q_r}{ S_q q_r+S_\tau }
\ee
that describes a  transformation of a rate of the DoF $q$ (e.g. co-ordinate)
with respect to the reference DoF $\tau$ under the change to the reference DoF $r$.
 - interpreted as an explicit manifestation of the {\bf 'time relativity'}.
\section{Conclusion}
In the present note, basic principles of dynamics and derivation of dynamical equations are revised and re-ordered in an alternative formulation.
 
\subsection{ Short summary of outstanding inconsistencies to be solved by an alternative approach }
\begin{itemize}
\item
{\bf  Time direction and reversibility - }\\
 inconsistency between the reversibilty of time in dynamical equations and the strict direction of irreversible evolution for complex systems

\item
{ \bf  Causality  - }\\
ambiguity of definition for  {\it cause - result } sequence
especially for relativistic and quantum systems 

\item
{ \bf  ambiguity of time definition - }
time flows  differently in different systems at different conditions
\end{itemize}

A possible way to remediation of these shortcomings could be a partial revision 
 of our understanding of the dynamics and first of all, the phenomena   of time appearing, measurement and interpretation.

These are the conclusions following from the time as defined by measurement: 
\bi
\item
there is no universal time as a physically well defined measure;
\item
 not stastes are ordered in time, but the time arises as a result of ordering;
\item
the basic measure for ordering is the entropy; thus the phonomen of "time" has a pure statistical issues;
\item 
 to fix the {\it local } evolution direction, an additional condition (conditions) - ergodicity- should be kept;
it is usually interpreted as a conservation law (laws);

\item
for  a system with more as 2 DoF's, the  first order gradient with only
one ergodicity condition provides not a unique equatioins: 
 
 to obtain a uniqueness there are two possibilities\\
 1. consider more ergodicity equations (conservations)\\
 2. extend the enropy variation to higher order gradients
\item
 the state evolution results form the principle of {entropy gradient maximization}
 in form of evolution equations; together with initial conditions is the evolution
 uniquely defined;

\item
 as a time measure, a state sequence of arbitrary subsytem (clock) can be taken and  is therefore related to this subsystem
 especially, the relativity of time is a manifestation of this relation;

\item
 the dynamic equations are the evolution equations with the separated DoF's
of the chosen clock system;

\item change of the time variable from one clock-subsystem to another implies
  a transformation of time 
\ei

 These conclusions are illustrated and  discussed on the known example  of classical hamiltonian systems 
 systems with discoupled time DoF,
 especially with a large number of DoF's (stastistical systems) subjected to Boltzmann dynamical equations.
\cite{hamiltonian}

 It is worth noting that some usual 	conceptions 
are conversed in the philosophy of the present formalism, such as:
\bi
\item
    solution of dynamical equations should obey the second thermodynamic law for global systems\\
$\ra $   the second thermodynamic law generates the  dynamical equations

\item
   the entropy increases in the direction of time\\
$\ra $    the direction of time is determined by increasing  entropy

\item
     conservation laws follow from the solutions of dynamical equations\\
$\ra $    the dynamical equations resut from  conservation laws interpreted as ergodicity conditions

\item
clock ticks differently in different reference systems, because of the time relativity \\
$\ra $  the time is defined by clock ticking.
\ei 

\subsection{  \textbf{  Technical  advances  of the formalism    }}     

\begin{itemize}
\item the formalism takes for a basis only one well justified governing principle

\item
 the formulation is strongly {\bf deterministic}:\\
  all possible states and the structure of the state space for a closed system are determined by  its DoF's completely\\
 
\item the formulation is strongly {\bf local}:\\
a next state is determined completely by local maximization and initial conditions respectivelythe trajectories are determined by initial points and the entropy gradient maximized under additional condition(conditions)

\item 
 thus the topological structure of the state space is determined by paths (trajectories) between 
a  starting point ( {\it cause } )-  the initial condition of path; the causality is then fixed uniquely;

\item
 'time dynamics' is the ordering of the points of trajectory with respect to substates of a chosen subsystem  (clock system) of the same closed system;

It means, that the time is only deined with respect the reference subsystem (chosen as a clock) and its degrees of freedom.
 For example, a time interval between two events of a microscopic system in quantum mechanics is defined in fact in a macroscopic system of observation.

 In a contrary, the term 'age of the Universe' is generally meaningless, since no reference system can be defined.
\end{itemize}

\subsection{ General   advances  resulting from the formalism  }
\vs{.01}
\begin{itemize}
\item
  the formalism contains  only one (just existent and well known!)
 basic principle of a statistical issue - the entropy maximization
 
\item 
the time reversibitly is eliminated, since the evolution obeys a first-order conditions

\item 
the causality is uniquely fixed as a solution of the forst -order problem with initial condition

\item 
  postulates of the special relativity (existence of a finite rate, transformation of time), consequently the special relatvity itself as well - need not to be postulated anymore since all these  principles follow automatically from the formalism
\end{itemize}

\end{document}